\documentclass[twocolumn,showpacs,aps]{revtex4}

\begin{document}

\preprint{Procs. of IV DGFM-SMF Workshop on Gravitation and
Mathematical Physics, November 2001, Chapala, M\'exico.}

\title{Is power-law inflation really attractive?}

\author{C\'esar A. Terrero-Escalante}
\email{cterrero@fis.cinvestav.mx}
\affiliation{Instituto~de~F\'{\i}sica,~UNAM,
~Apdo.~Postal~20-364,~01000,~M\'exico~D.F.,~M\'exico.}%


\begin{abstract}
It is argued that the order of the analytic expressions for the
calculation of the primordial perturbations from inflation exerts
a strong influence upon the results of the analysis of observables
dynamics based on these expressions and, therefore, upon some
predictions sometimes taken for granted as generic for the
inflationary scenario.
\end{abstract}

\pacs{PACS numbers: 98.80.Cq, 98.80.Es, 98.70.Vc}
\maketitle

\section{\label{sec:Intro}Preliminaries\protect}

The idea about the energy density in the very early universe being
dominated by the potential energy of a single real scalar field,
commonly called inflaton, is strongly supported by the analysis of
recent cosmological observations \cite{inflation,CMBdata}. In this
scenario density and metrics quantum fluctuations were stretched
beyond the causal horizon due to the accelerated expansion
produced by a negative pressure. Much after the inflationary
period is ended these fluctuations reentered the causal horizon
giving rise to perturbations in the gravitational potential with
an almost scale-invariant spectrum. Depending on the exact time of
reentering the causal universe these perturbations became the
seeds for anisotropies in the cosmic microwave background (CMB)
radiation or for the large scale structure formation. Therefore,
measuring the spectra of CMB anisotropies and density distribution
in the observable universe, the corresponding spectrum of
primordial perturbations can be determined. Each inflaton
potential characterizes different physics and, consequently,
different spectra of fluctuations. This way, determining the
primordial spectra can give us important hints about the physics
in the very early universe.

Density (scalar) and metrics (tensorial) perturbations can be
described by means of the spectral indices \cite{inflation},
\begin{eqnarray}
\label{eq:nSDef}
\Delta&\equiv&\frac{n_S-1}2 \equiv \frac{d\ln A_S}{d\ln k} \, ,\\
\label{eq:nTDef} \delta&\equiv&\frac{n_T}2 \equiv \frac{d\ln
A_T}{d\ln k} \, ,
\end{eqnarray}
where the subscripts $S$ and $T$ stand for scalar and tensor modes
respectively, $A_{S,T}$ denotes the corresponding normalized
spectral amplitudes and $k=aH$ is the comoving wavenumber when the
mode crosses the Hubble radius, $d_H\equiv H^{-1}\equiv
a/\dot{a}$, during inflation. Here $a$ is the scale factor of the
universe spatial volume and dot refers to derivatives with respect
to cosmic time, $t$.

One of the very few models where closed analytic expressions of
the perturbation spectra are available is power-law inflation,
$a(t) \propto t^p$ where $p>1$ is a constant \cite{PLinfl}. For
this model the Hubble parameter, $H$, and the potential, $V$, as
functions of the inflaton field, $\phi$, are given by,
$H\propto\exp(-\sqrt{\kappa/2p}\,\phi)$ and
$V\propto\exp(-\sqrt{2\kappa/p}\,\phi)$, where $\kappa =
8\pi/m_{\rm Pl}^2$ is the
 Einstein constant and $m_{\rm Pl}$ is the Planck mass.
This model has several attractive features concerning theory and
observations. First of all, the exponential function is the limit
of a Taylor series similar to the tree expansion often used for
scalar field potentials arising in quantum field theory
\cite{inflation}. Second, the exponential potential is the only
inflationary model yielding exact power-law spectra \cite{abbott},
$A_S\propto k^{\Delta}$ and $A_T\propto k^{\delta}$. The deviation
of these spectra amplitudes from scale-invariance (also known as
the tilt of the spectra) is given by the constant spectral indices
$\delta=\Delta= -(1/p+1/p^2+1/p^3 +\cdots) \leq 0$. Since the
spectra scale-dependence is predicted to be very weak, the
power-law parametrization of the primordial spectra is used very
often while analyzing the CMB data \cite{CMBdata}.

For more general potentials the spectra must be calculated
numerically or by means of approximations based on the smallness
of the horizon flow functions \cite{inflation,Schwarz:2001vv},
\begin{equation}
\label{eq:Hjf} \epsilon_{m+1} \equiv \frac{{\rm d} \ln |\epsilon_m|}
{{\rm d} N}, \qquad m\geq 0 \, ,
\end{equation}
and,
\begin{equation}
\epsilon_0 \equiv \frac{{d_{\rm H}}(N)}{d_{\rm Hi}},
\end{equation}
where $N \equiv \ln (a/a_{\rm i})$ is the number of e-folds since
some initial time $t_{\rm i}$, and $d_{\rm Hi} \equiv d_{\rm
H}(t_{\rm i})$. According to (\ref{eq:Hjf}),
\begin{equation}
\epsilon_1 \equiv \frac{{\rm d} \ln d_{\rm H}}{{\rm d} N} \, .
\end{equation}
Inflation happens for $\epsilon_1 < 1$ (equivalent to $\ddot{a} >
0$) and $\epsilon_1 >0$ from the weak energy condition (for a
spatially flat universe). For $m>1$, $\epsilon_m$ may take any
real value. Expressions (\ref{eq:Hjf}) define a flow in the space
$\{ \epsilon_m\}$. This flow is described by the equations of
motion
\begin{equation}
\label{eq:eom} \epsilon_0\dot{\epsilon_m} = \frac1{d_{\rm
Hi}}\epsilon_m \epsilon_{m+1} \, .
\end{equation}

According with Eqs.~(\ref{eq:eom}), for power-law inflation we
have $\epsilon_1=1/p=$ constant and $\epsilon_m=0$ for $m>1$.

So far, in the case of a slowly rolling inflaton, the more precise
expressions for the spectral amplitudes were derived in
Ref.~\cite{SG}. (Rigourously speaking, the scalar amplitudes were
derived by Stewart and Gong \cite{SG}. Using the same procedure
the tensorial ones were derived later by Leach et al.
\cite{Leach:2002ar}). From these next-to-next-to-leading order
expressions for the amplitudes and using definitions
(\ref{eq:nSDef}) and (\ref{eq:nTDef}), the spectral indices are
obtained,
\begin{eqnarray}
\label{eq:SLns} \Delta= &-&\epsilon_1 - \frac{1}{2}\epsilon_2 -
\epsilon_1^2 - (C+\frac{3}{2})\epsilon_1\epsilon_2 -
\frac{C}{2}\epsilon_2\epsilon_3
\nonumber \\
&-&\epsilon_1^3 - (3C - \frac{\pi^2}{2} +
\frac{17}{2})\epsilon_1^2\epsilon_2 \nonumber \\
&-& (\frac32C + \frac{C^2}2 - 7\frac{\pi^2}{24} + \frac72)\epsilon_1\epsilon_2^2 \nonumber \\
&-& (2C + \frac{C^2}2 - 7\frac{\pi^2}{24} +
\frac72)\epsilon_1\epsilon_2\epsilon_3 \nonumber \\
&-& (1-\frac{\pi^2}8)\epsilon_2^2\epsilon_3 - (\frac{C^2}4 -
\frac{\pi^2}{48})\epsilon_2\epsilon_3^2 \nonumber\\
&-& (\frac{C^2}4 -\frac{\pi^2}{48})\epsilon_2\epsilon_3\epsilon_4
\, , \\ \nonumber
\label{eq:SLnt}
\delta = &-&\epsilon_1 -
\epsilon_1^2 - (C+1)\epsilon_1\epsilon_2 \\ \nonumber &-&
\epsilon_1^3 -(3C-\frac{\pi^2}{2}+8)\epsilon_1^2\epsilon_2 \\
\nonumber
&-&(\frac{C^2}{2}+C-\frac{\pi^2}{24}+1)\epsilon_1\epsilon_2^2
\\
&-&(\frac{C^2}{2}+C-\frac{\pi^2}{24}+1)\epsilon_1\epsilon_2\epsilon_3
\,.
\end{eqnarray}

It is easy to check that the limit of a very slowly rolling
inflaton ($\epsilon_m\rightarrow 0$) coincides with power-law
inflation. Since the current precision of the CMB data does not
allow to discern any scale dependence for the spectral indices, it
is commonly assumed that the underlying inflationary scenario must
belong to the class of extreme slow-roll inflation. In this
contribution it is shown how the qualitative analysis of the
inflationary dynamics as described by the expressions used to
calculate the primordial spectra strongly depends on the order of
these expressions.

\section{\label{sec:disc}Discussion}


The leading order is recovered from Eqs.~(\ref{eq:SLns}) and
(\ref{eq:SLnt}) by neglecting terms with order higher than the
linear one,
\begin{eqnarray}
\label{eq:SLnslo} \Delta= &-&\epsilon_1 - \frac{\epsilon_2}{2}\,,\\
\label{eq:SLntlo} \delta = &-&\epsilon_1 \,.
\end{eqnarray}
First of all, let us note that strictly speaking and according
with Eqs.(\ref{eq:eom}), neglecting terms like
$\epsilon_1\epsilon_2$ implies $\epsilon_1=$ constant and
$\epsilon_m=0$, leading to power-law inflation. However, in our
approximation we will allow quadratic terms being neglected
without necessarily neglecting any linear term. In other words,
the horizon flow functions are assumed to be very weakly
time-dependent. Using definitions (\ref{eq:Hjf}) we can rewrite
system (\ref{eq:SLnslo}) and (\ref{eq:SLntlo}) as,
\begin{equation}
\label{eq:lo} \hat{\delta}=\delta-\Delta\, ,
\end{equation}
where a circumflex accent denotes differentiation with respect to
the variable $\tau$, defined such that $d\tau\equiv d\ln H^2$ and
$d\tau/dN=-2\epsilon_1$. As can be noted, power-law inflation is a
trivial solution to this equation and, in fact, it is a fixed
point of the dynamics described by this equation. Since this is a
first order differential equation this fixed point can only be an
attractor or a repellor. We have two observables we would like to
trace for in order to make generic predictions for inflation,
namely, $\delta$ and $\Delta$. Because the differentiation only
involves the tensorial index, we are forced to make assumptions on
the behavior of the scalar index or, at least, on the difference
between the scalar and tensorial indices, taking into account that
this difference is generally a function of time. Conditions upon
$(\delta-\Delta)$ can be translated into conditions upon
$\epsilon_2$ which in terms of the inflaton potential and its
derivatives with respect to $\phi$ (denoted with prime) reads
\cite{Leach:2002ar},
\begin{equation}
\label{eq:e2} \epsilon_2 \approx
\frac{2}{\kappa}\left[\left(\frac{V'}{V}\right)^2-\frac{V''}{V}\right]\,
.
\end{equation}
Hence, power-law inflation will be an attractor of the dynamics
described by Eq.~(\ref{eq:lo}) only if $\epsilon_2>0$, i.e., if
$V'^2>VV''$ (note that $d\tau/dt<0$). This will be the case for
any potential with $V''<0$ (in this paper we are considering the
inflaton potential to be nonnegative). Examples of models with
this kind of potential are the inverted quadratic model, the
cosinus model, models with $V\propto\phi^{\frac nm}$ where $n<m$
are some integers, etc. On the other hand, there are models with
$V''>0$ and satisfying $V'^2>VV''$. For instance, we have all the
monomial models with $V\propto\phi^p$ and $p>1$ a real number.
Indeed, the above listed models are amongst the most popular
inflationary scenarios \cite{inflation}, however, it is possible
to find some other interesting models with $V''>0$ but such that
$V'^2<VV''$, i.e., scenarios where power-law inflation will be a
repellor rather than an attractor. One of these models is, for
instance, the hyperbolic cosinus having Taylor series quite close
to the scalar field potential tree expansion with only even order
terms \cite{inflation}. This model provides a counterexample to
the leading order analysis by Hoffmann and Turner \cite{HT} which
yields power-law inflation as a generic attractor of the
inflationary dynamics. The point here is that $x''$, where
$x\equiv V'/V$, assumed in Ref.~\cite{HT} to be constant, for the
hyperbolic cosinus model is equal to $-2a^3sinh(ax)/cosh^3(ax)$,
and will not be approximately constant unless $a\ll 1$.

For the next-to-leading order analysis we must keep only up to
second order terms in Eqs.~(\ref{eq:SLns}) and (\ref{eq:SLnt}).
After converting to differential equations in terms of $\tau$
\cite{Ayon-Beato:2000xx},
\begin{eqnarray}
\label{eq:SLnsnlo} 2C\epsilon_1
\hat{\hat{\epsilon_1}}-(2C+3)\epsilon_1 \hat{\epsilon_1}
-\hat{\epsilon_1}
+\epsilon_1^{2}+\epsilon_1 +\Delta &=&0\,,  \\
\label{eq:SLntnlo}
2(C+1)\epsilon_1\hat{\epsilon_1}-\epsilon_1^{2}-\epsilon_1 -\delta
&=&0\,,
\end{eqnarray}
where $C=-0.7296$. Let us note here that this system of
differential equations it is not a closed system. Even if we
differentiate Eq.~(\ref{eq:SLntnlo}) with respect to $\tau$ and
substitute the result and Eq.~(\ref{eq:SLntnlo}) itself in
Eq.~(\ref{eq:SLnsnlo}), we still have three independent variables
to deal with. Information on the functional forms of one of the
observables $\Delta$ and $\delta$ is needed again to describe the
dynamics of $\epsilon_1$
\cite{Ayon-Beato:2000xx,Ayon-Beato:2000ga,Terrero-Escalante:2001ni}.

The dynamical analysis of Ref.~\cite{Ayon-Beato:2000xx} indicates
that in the reduced phase spaces ($\Delta=$ constant) for the
evolution of $\epsilon_1$ as described by Eq.~(\ref{eq:SLnsnlo}),
there exists a saddle point in the region where $\epsilon_1$ has
interesting values and $\Delta<0$. Thus, with respect to cosmic
time, attractor--like behavior will be characteristic only of
those trajectories that are very close to the unstable
separatrices. Likewise, the saddle point acts as a repellor for
those trajectories that are closer to the stable separatrices. For
blue tilted spectra, $\Delta>0$, there is not fixed point of any
kind in the physically interesting range of $\epsilon_1$
\cite{Ayon-Beato:2000xx}. The first order constrain given by
Eq.~(\ref{eq:SLntnlo}) seems to force the dynamics of $\epsilon_1$
to do not differ too much from power-law inflation if the spectral
indices are to be weakly scale-dependent
\cite{Terrero-Escalante:2001ni,Terrero-Escalante:2001rt}.
Inflationary dynamics that, to next-to-leading order, do not
correspond to a power law attractor were found in
Refs.~\cite{Terrero-Escalante:2001rt,Terrero-Escalante:2001du},
though a transient regime of power-law behavior exists in both
cases. These models provide further counterexamples to the leading
order results in Ref.~\cite{HT}.

To next-to-next-to-leading order the system of differential
equations corresponding to Eqs.~(\ref{eq:SLns}) and
(\ref{eq:SLnt}) is,
\begin{eqnarray}
\label{eq:SLnsnnlo} &-&
(2C^2-\frac{\pi^2}{12})\epsilon_1\hat{\hat{\hat{\epsilon_1}}}
\nonumber \\
&-&
(2C^2-\frac{7\pi^2}{12}+8)\epsilon_1\hat{\epsilon_1}\hat{\hat{\epsilon_1}}
+ (2C^2+8C-\frac{7\pi^2}{6}+14)\epsilon_1^2\hat{\hat{\epsilon_1}}
\nonumber \\
&+& (2C^2+6C-\frac{7\pi^2}{6}+14)\epsilon_1\hat{\epsilon_1}^2 -
(6C-\pi^2+17)\epsilon_1^2\hat{\epsilon_1} \nonumber \\
&+& 2C\epsilon_1
 \hat{\hat{\epsilon_1}}-(2C+3)\epsilon_1 \hat{\epsilon_1}
\!-\!\hat{\epsilon_1}\! + \! \epsilon_1^{3} \! + \! \epsilon_1^{2}
\! + \! \epsilon_1 \!+ \!\Delta =0\,,
\end{eqnarray}
and,
\begin{eqnarray} \label{eq:SLntnnlo} &&
(-2C^2-4C+\frac{\pi^2}{6}-4)\epsilon_1^2\hat{\hat{\epsilon_1}}
\nonumber \\
&+&
(-2C^2-4C+\frac{\pi^2}{6}-4)\epsilon_1^2\hat{\epsilon_1}\hat{\epsilon_1}^2
-(-6C+\pi^2-16)\epsilon_1^2\hat{\epsilon_1}\nonumber \\
&+&
2(C+1)\epsilon_1\hat{\epsilon_1}-\epsilon_1^3-\epsilon_1^{2}-\epsilon_1
-\delta =0\,,
\end{eqnarray}

The complete qualitative analysis of this dynamical system will be
carried out in a forthcoming publication, let us, however, note
that now, the dynamics described by the reduced ($\delta=$
constant) second order equation (\ref{eq:SLntnnlo}) is very
similar to that described by the reduced ($\Delta=$ constant)
equation (\ref{eq:SLnsnlo}) which was fully analyzed in
Ref.~\cite{Ayon-Beato:2000xx}. In particular, the eigenvalues of
the corresponding Jacobian matrix have the form
$\lambda_{\pm}=(g_2\pm \sqrt{g_2^2+4g_1})/2$, where $g_{1,2}$ are
functions of the roots of equation
$\epsilon_1^3+\epsilon_1^2+\epsilon_1+\delta=0$. It can be shown
that for $\delta<0$ the eigenvalues will have opposite sign
indicating the existence in the reduced phase space of a saddle
point corresponding to $\epsilon_1=$ constant. Thus, to
next-to-next-to-leading order there is not more a constrain
forcing the dynamics to be close to power-law inflation and this
scenario becomes less attractive than to lower orders.

\section{\label{sec:conc} Conclusions}

It was argued that, even if the power-law inflationary model has
very attractive features from the theoretical and observational
points of view, and it is the limit of scenarios with extremely
slow rolling inflaton fields, it cannot be claimed that such a
scenario must be expected to be an attractor of the inflationary
dynamics driven by general potentials, particularly those that do
not belong to the class describing a very slow rolling regime. In
general, any claim about generic predictions for the inflationary
observables dynamics must take into account the order of the
expressions involved unless a non-dependence on the order is
granted. This point concerns also the reliability of programmes
for the inflaton potential reconstruction. The order to be used to
calculate the primordial spectra will be fixed by the quality of
the forthcoming cosmological observations, in particular by the
capability of detecting any scale dependence of these spectra.
\begin{acknowledgments}
Supported in part by the CONACyT grant 38495--E.
\end{acknowledgments}


\end{document}